\documentclass[letter]{spie} 

\usepackage[T1]{fontenc}
\usepackage[english]{babel}
\usepackage[utf8]{inputenc}

\usepackage{amsmath,amsfonts,amssymb}
\usepackage{array}
\usepackage{mathtools}
\usepackage{graphicx}
\usepackage{textcomp}
\usepackage{xcolor}
\usepackage[super]{nth}

\graphicspath{{figures/}} 
\usepackage[colorlinks=true, allcolors=blue]{hyperref}
\usepackage{bm}

\title{Controlling petals using fringes: discontinuous wavefront sensing through sparse aperture interferometry at Subaru/SCExAO}

\author[a$\star$]{V.~Deo}
\author[a,b]{S.~Vievard}
\author[c]{N.~Cvetojevic}
\author[a]{K.~Ahn}
\author[d]{E.~Huby}
\author[a,b,e,f]{O.~Guyon}
\author[d]{S.~Lacour}
\author[a]{J.~Lozi}
\author[c]{F.~Martinache}
\author[g,h,i]{B.~Norris}
\author[a,d,j]{N.~Skaf}
\author[g,h,i]{P.~Tuthill}

\affil[a]{\small Subaru Telescope, National Astronomical Observatory of Japan, National Institute of Natural Sciences, Hilo, HI, U.S.A.}
\affil[b]{Astrobiology Center of NINS, 2-21-1, Osawa, Mitaka, Tokyo, Japan}
\affil[c]{Université Côte d'Azur, Observatoire de la Côte d'Azur, CNRS, Laboratoire Lagrange, France}
\affil[d]{LESIA, Observatoire de Paris, Univ. PSL, CNRS, Sorbonne Univ., Univ. de Paris, 92195 Meudon, France}
\affil[e]{Steward Observatory, University of Arizona, Tucson, AZ, U.S.A.}
\affil[f]{College of Optical Sciences, University of Arizona, Tucson, AZ, U.S.A.}
\affil[g]{Sydney Institute for Astronomy, School of Physics, The University of Sydney, NSW 2006, Australia}
\affil[h]{Sydney Astrophotonic Instrumentation Laboratories, Physics Road, University of Sydney, NSW 2006, Australia}
\affil[i]{Astralis-USyd, School of Physics, University of Sydney 2006}
\affil[j]{Department of Physics and Astronomy, University College London, London, United Kingdom}

\authorinfo{$\star$ Corresponding author. Email: \texttt{vdeo} at \texttt{naoj} dot \texttt{org}.}

\begin{document}
\maketitle

\begin{abstract}
	Low wind and petaling effects, caused by the discontinuous apertures of telescopes, are poorly corrected -- if at all -- by commonly used workhorse wavefront sensors (WFSs).
	Wavefront petaling breaks the coherence of the point spread function core, splitting it into several side lobes, dramatically shutting off scientific throughput.
	We demonstrate the re-purposing of non-redundant sparse aperture masking (SAM) interferometers into low-order WFSs complementing the high-order pyramid WFS, on the SCExAO experimental platform at Subaru Telescope.
	The SAM far-field interferograms formed from a 7-hole mask are used for direct retrieval of petaling aberrations, which are almost invisible to the main AO loop.
	We implement a visible light dual-band SAM mode, using two disjoint 25 nm wide channels, that we recombine to overcome the one-lambda ambiguity of fringe-tracking techniques.
	This enables a control over petaling with sufficient capture range yet 	without conflicting with coronagraphic modes in the near-infrared.
	We present on-sky engineering results demonstrating that the design is able to measure petaling well beyond the range of a single-wavelength equivalent design.
\end{abstract}


			\section{INTRODUCTION}
			\label{sec:intro}

				\subsection{Low wind effect and petaling}
				\label{sec:intro:lwe}

Large telescopes with segmented apertures are prone to the nefarious low wind effect (LWE), also named island effect or petaling, which dramatically affects point spread function (PSF) quality.
LWE has been studied in great detail from VLT/SPHERE data\cite{Milli2018LWE}, and its occurrence is fairly common at Subaru Telescope as well\cite{Vievard2019LWEonSCExAO}.

\begin{figure}[t] 
	\centering
	{\footnotesize\bfseries
	VAMPIRES visible imager examples\\
	\includegraphics[width=.45\textwidth]{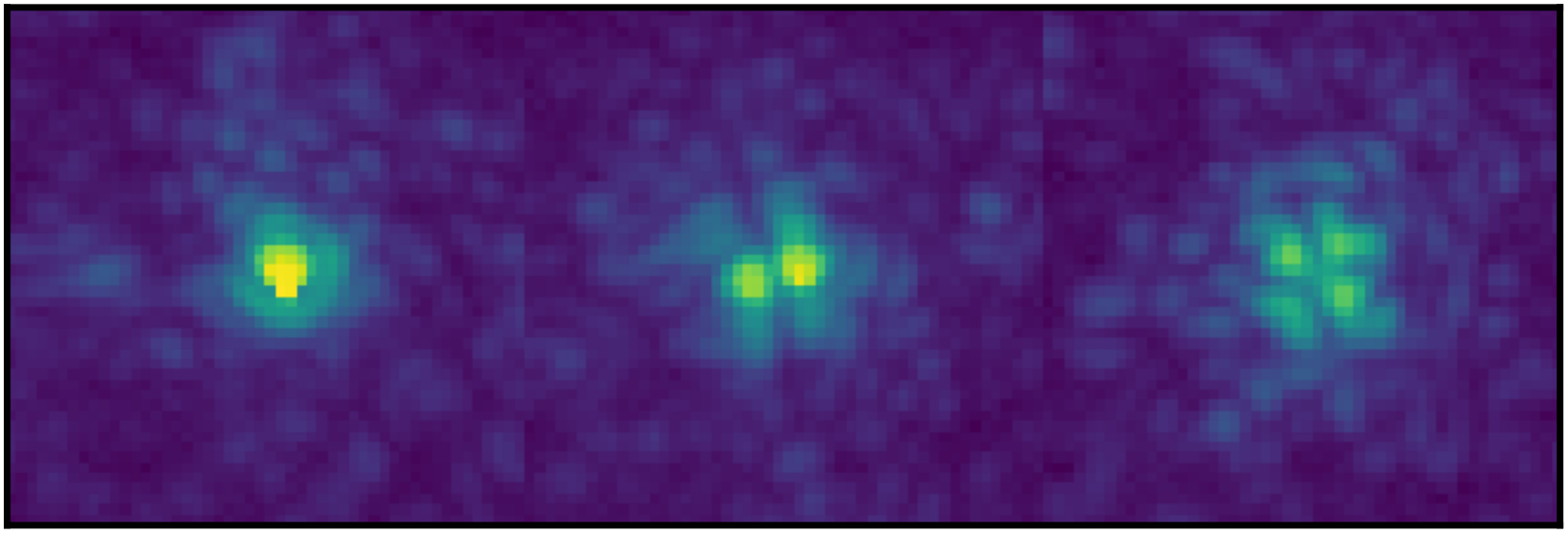}\\[1em]
	CHARIS integral field unit spectral cube
	} 
	\includegraphics[width=.9\textwidth]{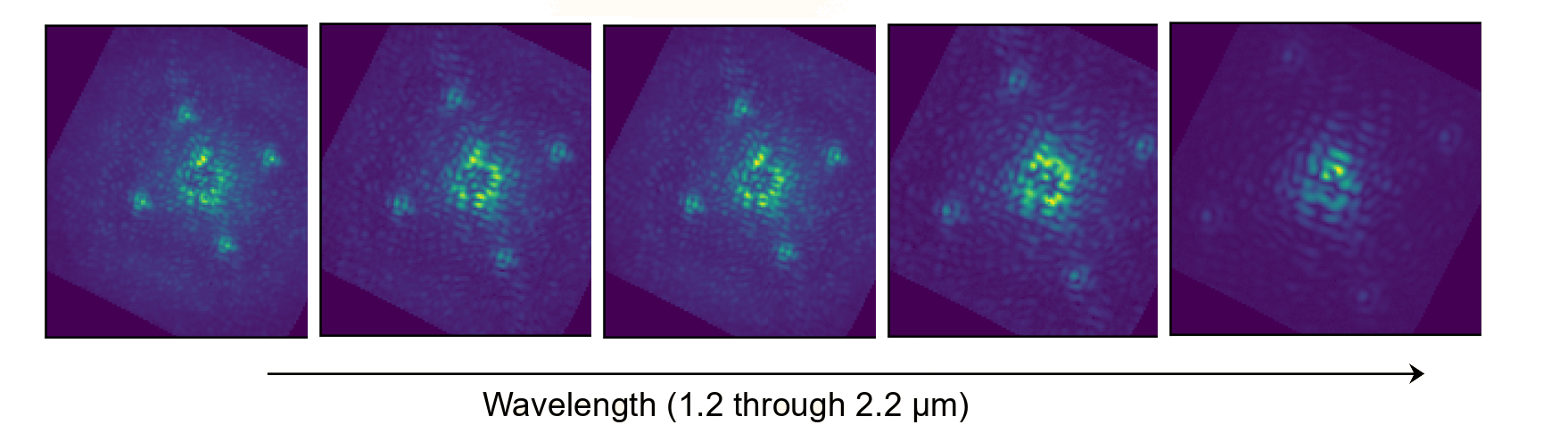}
	\caption{%
		Top: three selected on-sky PSFs acquired on Vega during Subaru/SCExAO engineering observations with the VAMPIRES visible imager (750~nm) during a night of excellent seeing (0.3''). 
		The left PSF is nominally diffraction limited, while the other two show LWE-induced core splitting.
		Bottom: samples of a CHARIS integral field unit spectral cube showing coronagraphic data from 1.2 to 2.2 \textmu m. Core splitting in the northeast-southwest direction behind the 0.113'' Lyot occulter increases the 0.2-0.3'' residual halo by several orders of magnitude.
		The four speckles in a square pattern within the dark hole area are astrometric calibration spots $15.2~\lambda/D$ from the optical axis.
	}
	\label{fig:LWE}
\end{figure}

We show on figure~\ref{fig:LWE} two examples of the imaging quality reduction induced by LWE on Subaru/SCExAO\cite{Currie2020OnSkySCExAO, Lozi2020SCExAOStatus} (more details about the instrument are provided in section~\ref{sec:exp_setup}).
First, we show core splitting in a fast imaging dataset using the VAMPIRES visible imager\cite{Norris2014Vampires, Uyama2020VampiresHAlpha}, where a significant fraction of the frames need to be discarded despite the seeing being an excellent 0.3''.
Second, we show the impact of core splitting on a coronagraphic spectral cube acquired with the CHARIS\cite{Lawson2020HD15115,Currie2021AcceleratingStars} integral field unit.
Just a few second-long bursts of core splitting during the exposure induce a significant brightening of the halo at separations within 0.3''.

\begin{figure}[b] 
	\centering
	\raisebox{2em}{\includegraphics[width=.15\textwidth, angle=-10]{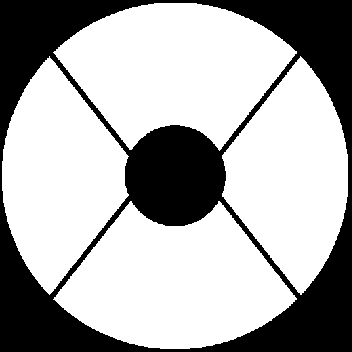}}
	\hspace*{.5em}
	\includegraphics[width=.7\textwidth]{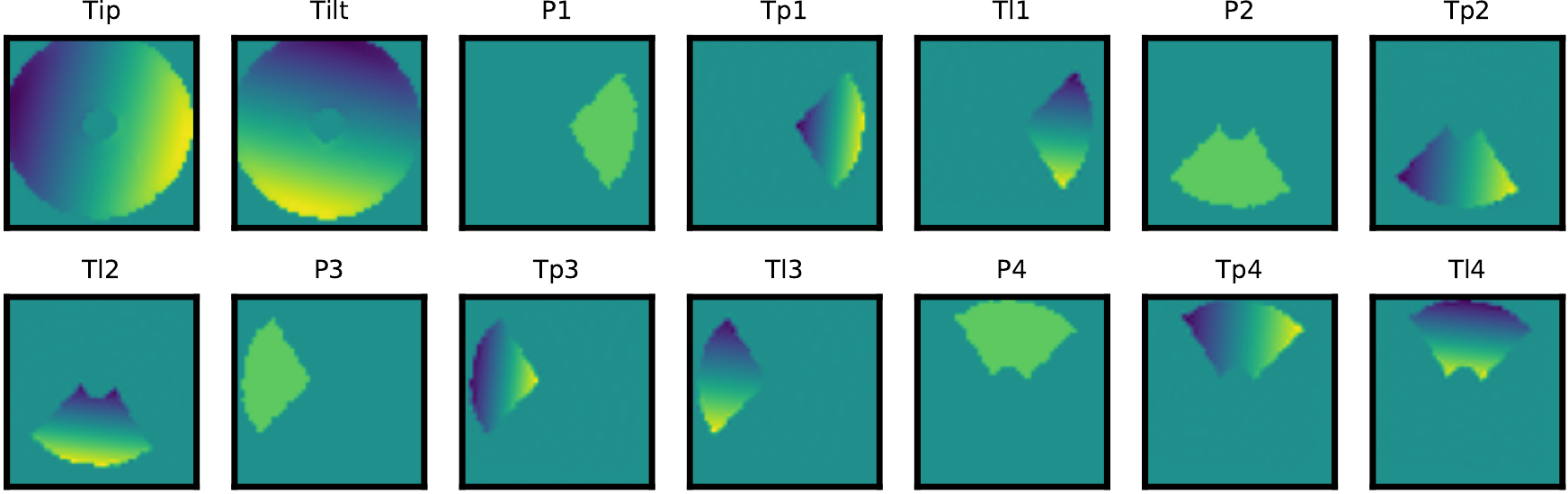}
	\caption{%
		Left: pupil of the Subaru Telescope (transmission map), showing the four petal components of the pupil.
		Right: wavefront maps for global tip-tilt and the 12 modes describing LWE for a 4-petal pupil: piston, tip, and tilt within each petal.
		These make for only 11 additional degrees of freedom since overall piston must be excluded.
	}
	\label{fig:TT_PTT_modes}
\end{figure}

The LWE aberration has been shown to be physically induced from thermal dissipation from the secondary support structure (spider) of the telescope, creating an optical path difference (OPD) jump between the sides of a spider arm.
This discontinuous aberration can be described fully, along with atmospheric turbulence, by typically extending a Kolmogorov-compatible wavefront mode basis with piston, tip, and tilt modes restricted to each individual clear subaperture (petal), as are shown on figure~\ref{fig:TT_PTT_modes}.
These additional modes make for 11 more degrees of freedom on a telescope with 4 spider arms.

				\subsection{Petal locking}
				\label{sec:intro:petallocking}
				
Traditional general purpose wavefront sensors (WFS) such as Shack-Hartmanns, Pyramids (PyWFS), or curvature WFSs, fare poorly for sensing a discontinuous wavefront step hidden under the shadow of the spider arm.
Adaptive optics (AO) systems may actually worsen petaling\cite{BertrouCantou2021Confusion} beyond what is physically introduced upstream by either physical LWE, or even just from natural seeing when the spider is larger than $\approx r_0$.
The AO can induce spurious stable wavefront control regimes where one or two of the four petals are stabilized one full wavelength (of the AO sensor) away from an appropriately flattened wavefront. We call this phenomenon petal locking.
As an example, Subaru/SCExAO uses a dual AO cascaded scheme.
The first stage AO188\cite{Minowa2010AO188} uses a 500-600~nm curvature WFS and drives a 188-element bimorph deformable mirror (DM); a wavefront discontinuity of 300~nm can be spuriously amplified and locked to $\approx$550~nm.
The second stage extreme AO within by SCExAO operates a 800-950~nm PyWFS and actuates a 50$\times$50 MEMS DM.
We observe petal steps of 600~nm created by AO188 being amplified to $\approx$900~nm by the extreme AO control loop.

Petal \emph{tip-tilt} modes are in theory not an issue for Shack-Hartmann or PyWFS sensors, which perform somewhat of a wavefront gradient measurement. 
Petal pistons, however, show zero gradient across the entire aperture, making their detection by PyWFSs extremely difficult, besides some diffractive effects yielding signal along the spider arms, and so only at lower modulation radii with of the order of 1 rad linear capture range\cite{BertrouCantou2020Petalometry}, while more than a full WFS wavelength of capture range will be necessary to recover from petal-locking situations.
By using an appropriate controller, a conventional AO will easily ``flatten'' the wavefront until only petal piston modes remain; which is why the focus of this paper is only the phasing of the three degrees of freedom of piston petals, and not the full additional 11 as shown on figure~\ref{fig:TT_PTT_modes}.

{}

Broadly speaking, we distinguish three different approaches in the AO community to tackle LWE/petaling at existing and future facilities.
(1) Eliminate discontinuous aberrations from the control space. This avoids petal-locking, but also removes capability to correct true, thermally induced LWE.
(2) Focus on developing petal-focused correction algorithms using signals from the main sensor.
This is a more accessible path for instruments designed with H or K band WFSs, where $r_0$ is larger than in the visible and expected discontinuities at the spider will remain within 1~rad.
And finally (3), develop an auxiliary WFS, with low resolution, dedicated to petaling.
In this paper, we propose to use non-redundant sparse aperture interferometry for petal wavefront sensing.

				\subsection{Sparse aperture non-redundant masking}
				\label{sec:intro:sawfs}
				
Sparse aperture masking (SAM) interferometry\cite{Tuthill2006SparseApAO,Tuthill2013NRMInterf} has been long used to circumvent the effects of atmospheric seeing, with the pupil shrunk to a sparse array of much smaller holes, over which wavefront error could be neglected. Having low speckle noise within each hole is a fundamental hypothesis of SAM. However, today a flat wavefront can be provided over holes significantly larger than $r_0$ thanks to (extreme) AO systems, allowing masks with a reduced number of larger holes.

\begin{figure}[b] 
	\centering
	\includegraphics[width=0.8\textwidth]{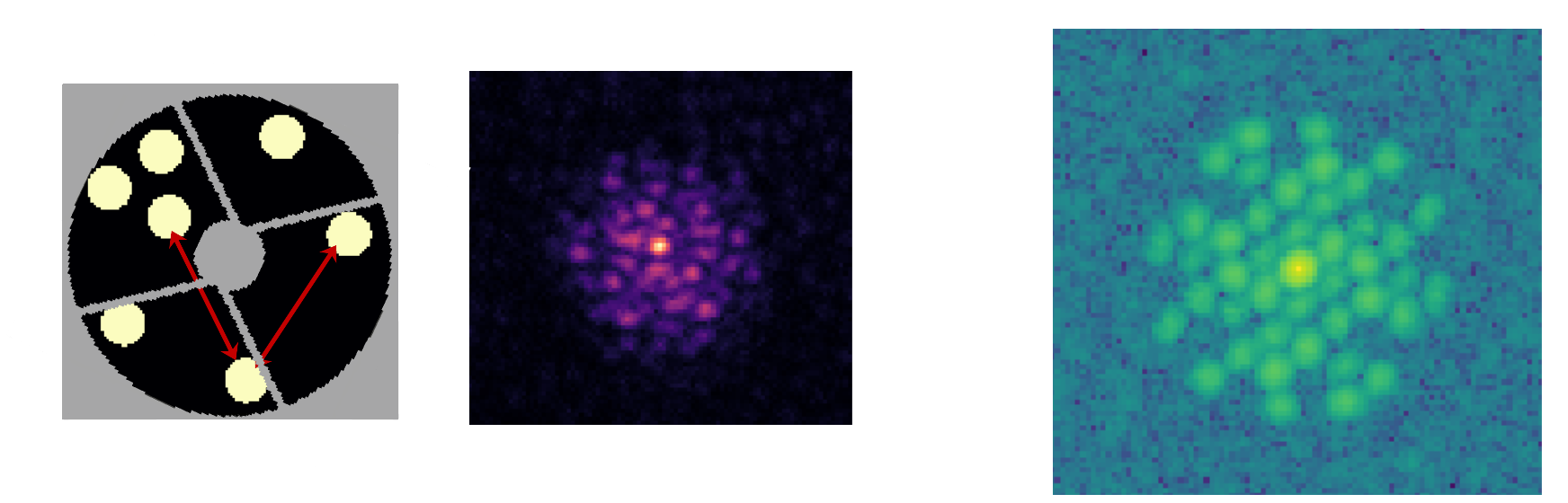}
	\caption{%
		Left: 7 hole non-redundant pupil mask, overlaid with the Subaru telescope pupil. Two of the 21 baselines are highlighted with red arrows.
		Middle: typical corresponding focal plane interferogram obtained on the VAMPIRES instrument at a wavelength of 775~nm.
		Right: complex magnitude of the corresponding correlogram, showing the 42 independent components corresponding two-to-one (each plus its Fourier symmetric) to the baselines of the 7 hole mask.
	}
	\label{fig:mask_ig_cg_example}
\end{figure}

After propagation from the SAM pupil to the focal plane, an interferogram (IG) is formed in place of the classical PSF, consisting of the superposition of interference fringes between each pair of holes of the pupil mask.
The Fourier transform of the IG is then the autocorrelation of the electric field in the pupil plane; we call these correlograms (CG).
We show on figure~\ref{fig:mask_ig_cg_example} the 7 hole mask, and typical on-sky interferogram and correlogram; these correspond to the setup used throughout this paper and detailed in section~\ref{sec:exp_setup}.

Under the condition that a non-redundant mask (NRM) is used, i.e. that no two baselines between holes are of the same length and orientation, the CG map is composed of disjoint domains distinctively identifying each baseline of the mask.
Within each of these domains, the complex argument of the correlogram depends only upon: (1) the structure of the imaged object, from which stems the scientific use of NRMs; as well as (2) the phase difference in the wavefront between the corresponding pair of holes.
The scientific component to the pair's complex argument is fixed, and equal to zero in the case of a point source.
Reading the complex argument of the correlogram offers a straightforward way to measure OPDs between all holes of the mask, and after reconstruction through a linear reconstructor, a low-order description of the wavefront, which in this case would be the phasing of the four petals of the pupil.

We therefore propose the study of using the sparse aperture masking concept for petal wavefront sensing, thus forming the Sparse Aperture Wavefront Sensor (SAWFS).
In previous work\cite{Deo2021SAWFS}, we explored the deployment of a SAWFS with a 9 hole NRM in the near infrared, and showed preliminary results of closed-loop petal piston control during on-sky engineering.
Several limitations arose from this design: (1) the transmission of the 9-hole NRM could be improved using a reduced number of larger holes. Indeed, only 4 holes are necessary for the petal-piston sensing problem at Subaru Telescope.
And (2), using a single IG in H-band restricted the capture range to half-wavelength steps, i.e. 800~nm, beyond which phase wrapping was unavoidable.
As discussed in section~\ref{sec:intro:petallocking}, 800~nm is an insufficient capture range for petal-locking effect in SCExAO.

In this paper, we expose the design of a SAWFS comprised of two cameras sensing IGs at 675 and 775~nm wavelength, both through the same 7~hole mask.
The transmission of the mask is improved to $15$\% of the full pupil, and the use of two wavelength permits the use of a phase unwrapping algorithm that 
extends the capture range well beyond steps of $\pm\lambda/2$, up to $\pm2.3$~\textmu m.
This provides enough range for petal-locking issues with red / near-infrared AO systems.
In section~\ref{sec:exp_setup} we detail the experimental hardware used for this experiment on Subaru/SCExAO; in section~\ref{sec:reduction}, we propose some elements of data reduction for single- and twin-wavelength SAWFS, in particular the phase disambiguation algorithm we use for two wavelengths.
In sect.~\ref{sec:lab}, we show experimental noise propagation results, and finally in \ref{sec:sky_ol}, we show the analysis of open-loop SAWFS measurements during an on-sky run.
The authors hope the readers have a pleasant time browsing through this paper.


			\section{SPARSE APERTURE WFS: EXPERIMENTAL SETUP}
			\label{sec:exp_setup}
					
The Subaru Coronagraphic extreme Adaptive Optics\footnote{%
	\url{http://www.naoj.org/Projects/SCEXAO/index.html}%
}\cite{Lozi2020SCExAOStatus,Currie2020OnSkySCExAO,Currie2021AcceleratingStars} (SCExAO) instrument is a versatile experimental high contrast imaging platform deployed at Subaru Telescope.
The core scientific purpose of SCExAO is to implement high-efficiency near-infrared coronagraphy, permitted by an extreme adaptive optics system, so as to achieve direct imaging of faint stellar companions\cite{Currie2020HD33632}, debris and protoplanetary disks\cite{Currie2017HD36546,Lawson2020HD15115,Steiger2021MECHIP109427}, and exoplanets\cite{Currie2022NatureABAur}.
A Strehl ratio of 80 to 90\% in H band is typically achieved in median seeing conditions at Maunakea observatories, up to 93-94\% for the best observing nights.

\begin{figure}[ht] 
	\centering
	\includegraphics[width=.99\textwidth]{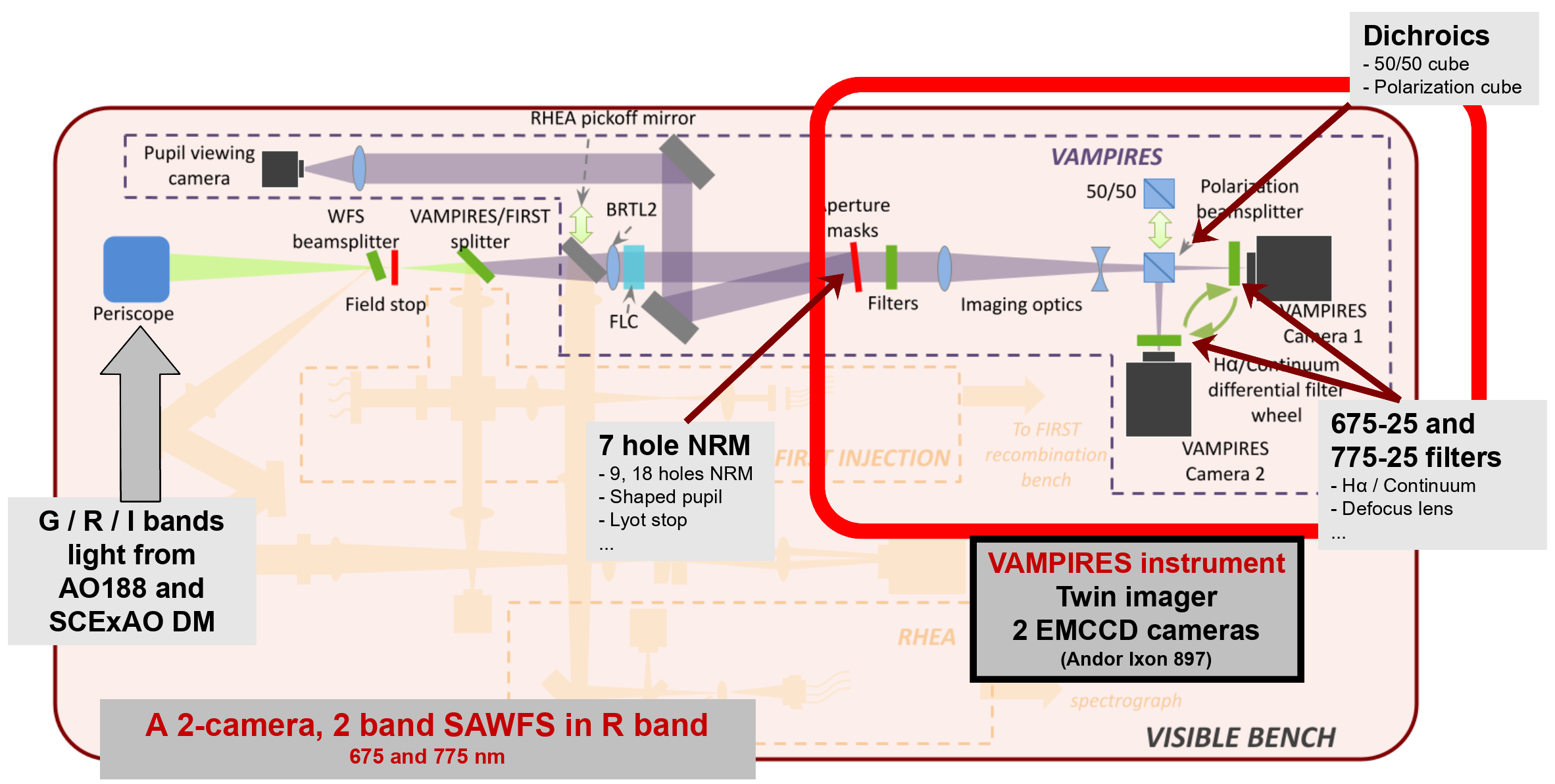}
	\caption{%
		Optical layout of the VAMPIRES instrument, with the key parts of the dual-band SAWFS, among the visible light bench of the SCExAO instrument.
		Not shown is the near-infrared part of SCExAO, which feeds light from 600 to 950~nm through the periscope (blue, top left), and contains the 50$\times$50 DM, and all scientific and technical modules operating in the 950~nm - 2.2~\textmu m wavelength range.
	}
	\label{fig:experimental_setup}
\end{figure}

Within SCExAO, the VAMPIRES \footnote{%
	Visible Aperture-Masking Polarimetric Interferometer/Imager for Resolving Exoplanetary Signatures
} instrument is a twin imager designed for differential polarimetric and H$\alpha$ imaging\cite{Norris2014Vampires}.
We show on figure~\ref{fig:experimental_setup} the layout of VAMPIRES within the upper bench of the SCExAO instrument.
Light from 600 to 950~nm is fed through the periscope from the lower deck of SCExAO, with the wavefront already corrected by both AO188 and the high resolution 2040~element DM of SCExAO.
I band is reserved for the extreme AO PyWFS, and the 600-800~nm range is fed into VAMPIRES.

VAMPIRES was recently expanded with coronagraphic capability\cite{Lucas2022VAMPIRESCoronagraph}, as well as with dual wavelength capability, using a dichroic splitter and two narrowband 675 and 775~nm filters.
The detectors are Andor Ixon 897 EMCCD cameras, which provide images with virtually no readout noise.
The total field of view is 3''$\times$3''; framerates of up to 360~Hz can be achieved using a reduced field of view of 0.37''$\times$0.37'', as we will use throughout this paper.

In order to enable analysis of stellar and circumstellar features below the diffraction limit, VAMPIRES was equipped with a suite of non-redundant masks, with respectively 7, 9, and 18 holes.
For the purpose of implementing a SAWFS, we focus on using the 7 hole NRM, which has the highest transmission of $\approx$15\% relative to the full pupil, while still having holes within each of the 4 pupil segments (figure~\ref{fig:mask_ig_cg_example}), thus allowing petal piston measurement.
It should be noted that having more holes that the absolute minimum of 4 required for petal sensing and control is both beneficial -- with the capability of simultaneous sensing, and thus discarding, tip-tilt from vibrations --, and suboptimal, since a lower number of holes allows greater overall transmission without overlap of baseline components in the CG plane.

			\section{ELEMENTS OF DATA REDUCTION}
			\label{sec:reduction}
				
				\subsection{Single camera SAWFS}
				\label{sec:reduction:single}
				
Briefly put, each component of the correlogram obtained from either camera (see figure~\ref{fig:mask_ig_cg_example}) relates to a single of the 21 baselines $ij$ between holes $i$ and $j$, where $0\leq i,j \leq 7$ are identifying indices of the 7 holes of the NRM.

The formalism and algorithmic process we use to reduce each camera to per-baseline measurements is described in much greater detail in reference~\citenum{Deo2021SAWFS}, with the same formalism as used here.
If $\mu_{ij}$ is the complex coherence over baseline $ij$ (obtained by sampling the CG at the peak of the corresponding sidelobe), the quantity of interest is its complex argument:
\begin{equation}
	\mathrm{Arg(\mu_{ij})} = \phi_i - \phi_j + \psi_{ij} +2k\pi,
\end{equation}
where $\phi_i$ is the value of the wavefront phase, at the sensing wavelength, within hole $i$ of the NRM, and $\psi_{ij}$ is the baseline's biasing phase, comprised both of an instrumental term and a physical component due to the imaged object not being a point source.
$k\in\mathbb{Z}$ represents the number of phase wraps between the actual phase difference and the measurement.

After an object-dependent calibration where reference phases $\psi_{ij}$ are determined, we define the phase measurement of the SAWFS as the collection for the 21 baselines, of the:
\begin{equation}
	m_{ij} = \mathrm{Arg}\left(\mu_{ij}\times\exp\left(-j\psi_{ij}\right)\right),
\end{equation}
which observes the relationship
\begin{equation}
	m_{ij} = \phi_i - \phi_j \text{ if} \left|\phi_i - \phi_j\right| \leq \pi.
\label{eq:m_ij}
\end{equation}
Equation~\ref{eq:m_ij} shows how we measure the wavefront phase difference between each pair of NRM holes, and shows that the capture range is limited to half a wavelength of OPD.
Pending that this condition is complied with on all 21 baselines, the wavefront can be reconstructed onto up to 6 modes using typically using a linear reconstruction:
\begin{equation}
	\label{eq:reconstruction}
	\mathbf{v} = \mathbf{D}^\dagger \bullet \dfrac{\lambda}{2\pi} \left[m_{ij}\right]_{0\leq i < j < 7},
\end{equation}
where $\mathbf{v}$ is the vector of modal command, $\mathbf{D}$ the response matrix from measurements (in units of OPD) to modes, $\bullet$ denotes matrix-vector product, and $\lambda$ the sensing wavelength.
				
				\subsection{Dual-wavelength recombination}
				\label{sec:reduction:dual}

Naturally, if some baseline phases $\phi_i-\phi_j$ exceed the $\pi$~rad capture range, the reconstruction $\mathbf{v}$ would be significantly wrong, and this could go undetected without additional checks on the SAWFS controller output.
This is why using two identical SAWFS at two different wavelengths $\lambda_0$ (=~775~nm) and $\lambda_1$ (=~675~nm) is essential to provide consistent measurements beyond half-wavelength phase jumps.

\begin{figure}[b] 
	\centering
	\includegraphics[width=.99\textwidth]{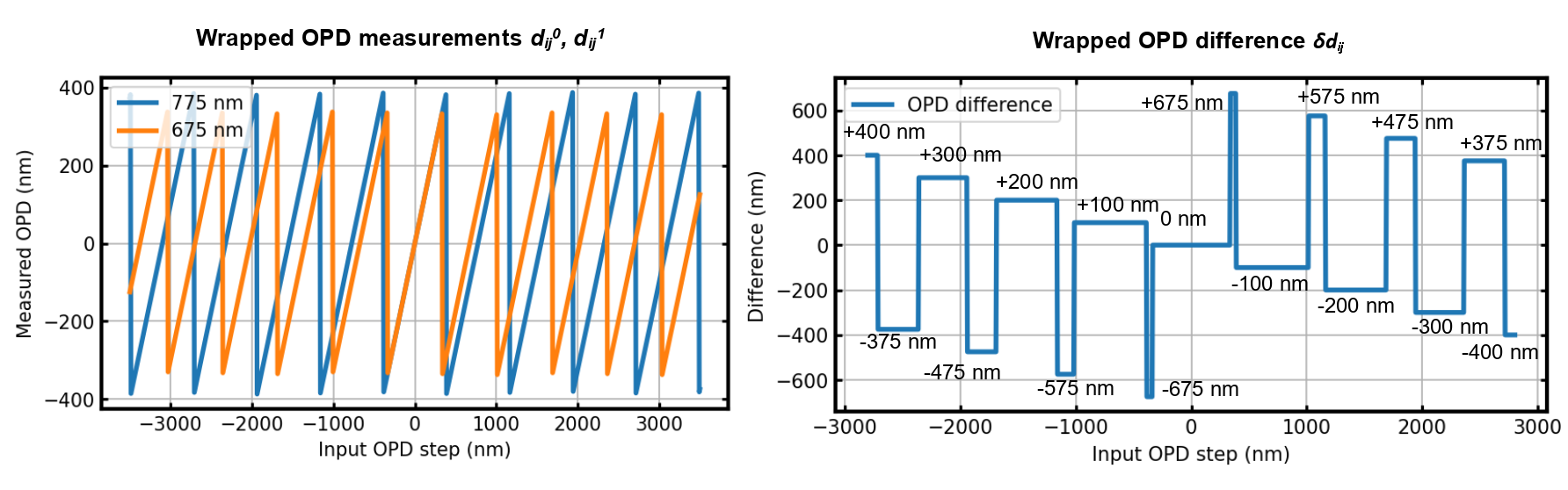}
	\caption{%
		Left: simulated wrapped OPD measurements $d_{ij}^0$ and $d_{ij}^1$ at both SAWFS sensing wavelengths for any baseline $ij$ as a function of the OPD difference between holes $i$ and $j$ of the NRM.
		Right: corresponding difference $\delta d_{ij}$, showing the discretization of the possible values used for retrieving the unwrapped OPD.
	}
	\label{fig:reduction:ddw}
\end{figure}

A fairly direct classification technique, that we call the differential dewrapper, is used to combine wrapped phase measurements from both cameras into a single, consolidated OPD measurement of much greater dynamic range.
Using the wrapped OPD quantity:
\begin{equation}
d_{ij}^{\{0, 1\}} = \dfrac{\lambda^{\{0, 1\}}}{2\pi}m^{\{0, 1\}}_{ij} = \dfrac{\lambda^{\{0, 1\}}}{2\pi} \left(\phi_i - \phi_j\right) + k\lambda^{\{0, 1\}},
\end{equation}
where the exponent $\cdot^{\{0, 1\}}$ denotes the index of the camera, we can compute the difference of wrapped OPDs between both spectral channels of the SAWFS:
\begin{equation}
\delta d_{ij} = d_{ij}^0 - d_{ij}^1,
\end{equation}
which, disregarding noise for now, can only take a discrete set of values.
This is illustrated on figure~\ref{fig:reduction:ddw} and allows applying the differential dewrapper.
The difference $\delta d_{ij}$ is related to the nearest permitted value, which maps to a number of phase wraps $k_0$ and $k^1$ for either wavelengths.
The unwrapped OPD is ultimately obtained by averaging both unwrapped values:
\begin{equation}
D_{ij} = \dfrac{\left(d_{ij}^0 - k^0\lambda^0\right) + \left(d_{ij}^1 - k^0\lambda^1\right)}{2},
\end{equation}
where both quantities on the numerator, if properly unwrapped and disregarding noise and differential biases between cameras, should be equal.

A key point to enable differential dewrapping to work, as can be observed from figure~\ref{fig:reduction:ddw}, is the closeness of the different possible discrete values for the wrapped OPD difference $\delta d_{ij}$.
For our chosen wavelengths of 675 and 775~nm, we observe that up to an input OPD of $\pm$2.36~\textmu m, the possible values for $\delta d_{ij}$ are all at least 100~nm apart from each other.
If we were to extend the capture range beyond 2.36~\textmu m, the next new possible values for $\delta d_{ij}$ are $\pm$375~nm, only at best 75~nm apart from the already allowed values $\pm 300$~nm.

As we will explore in section~\ref{sec:noise}, this specification on the classifier for $\delta d_{ij}$ gives a trade-off between acceptable noise and dynamic range. If we accept values only 75~nm apart, the range is extended to $\pm$2.71~\textmu m, at which point a difference occurs only 25~nm from an already existing one. With a tolerance of 25~nm, the range would be extended to $\pm$10.46~\textmu m, at which point $\delta d_{ij}$ start reoccurring again\footnote{
	Noting that 775 = 31 * 25, 675 = 27 * 25, that 27 and 31 are coprime, and 27 * 31 * 25~nm / 2. = 10462.5~nm, the mathematically expected half-period for re-occuring values $\delta d_{ij}$.
}.

We therefore arbitrarily define that with this dual wavelength configuration, the capture range of the SAWFS is $\pm$2.36~\textmu m, and that the noise specification in this case is for less than 50~nm error on all $\delta d_{ij}$, so as to avoid misclassification between possible values 100~nm apart.
Assuming we would want 50~nm of error to be a 2-sigma limit on $\delta d_{ij}$, this gives a standard deviation of 25~nm RMS, which, coming from both cameras, allows only a restrictive error budget of 18~nm RMS, i.e. $\approx 0.15$~rad per baseline per camera!

While we showed in previous work\cite{Deo2021SAWFS} that a single camera SAWFS could produce measurements will less that 1~rad RMS noise with only a few hundred photons --thanks to the high efficiency of interferometric measurements--, a 1~rad RMS error is excessive to allow the differential dewrapper to work.

In order to validate our actual capability to extend the dynamic range by using combined dewrapping of phases with both cameras,
we study in section~\ref{sec:noise} the noise propagation over a single camera; aftewards, we study in section~\ref{sec:sky_ol:results_noise} the a posteriori data from an on-sky dataset to analyze whether the noise conditions were met.


			\section{Experimental noise sensitivity}
			\label{sec:noise}
					
				\subsection{Photon-to-noise transfer on each camera, each baseline}

\begin{figure}[htbp] 
	\centering
	\includegraphics[width=0.8\textwidth]{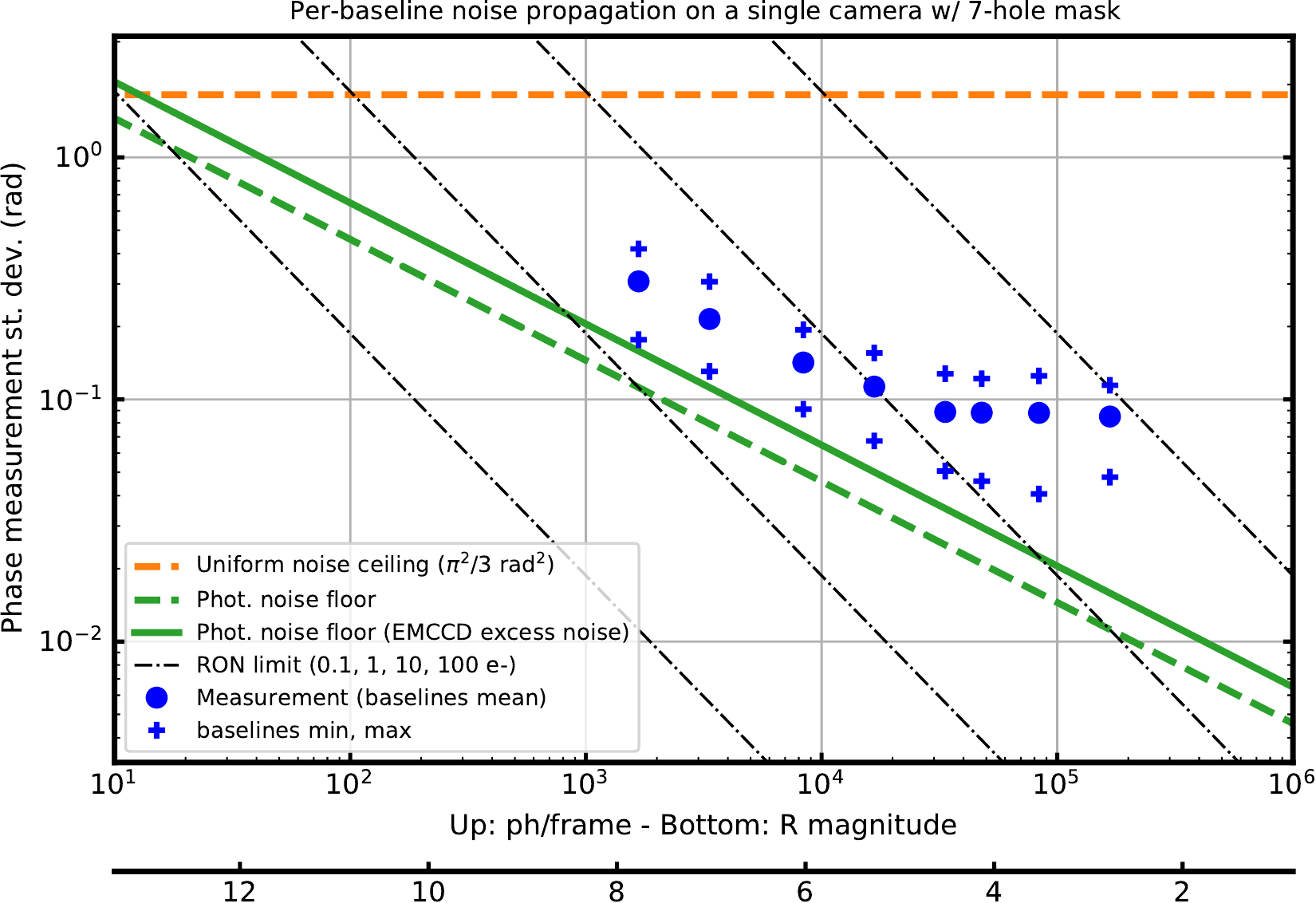}
	\caption{%
		Simulated and experimental noise propagation for a single-wavelength SAWFS using our 7 hole NRM. Propagation is measured from the number of photons per frame onto the phase noise \emph{per baseline}. Corresponding R-band magnitudes are given using the VAMPIRES zero-point of $8.24\ 10^7$ ph.frame\textsuperscript{-1}.second\textsuperscript{-1}.nm\textsuperscript{-1} per camera.
		Shown are the simulated photon noise limit without and with $\sqrt{2}$ excess noise; the ceiling corresponding to a uniform phase distribution over $[-\pi, \pi]$; and the read-out-noise floors for 0.1, 1, 10, and 100 $e^-$/px.
		Blue markers show experimental data: the mean and extrema across the 27 baselines of the NRM at the given photon fluxes, from $1.67\,10^3$ through $1.67\,10^5$ photons/frame.
	}
	\label{fig:noise_propagation_chart}
\end{figure}

Using the calibration source of the SCExAO bench, we analyze experimentally the noise propagation on the baseline phase measurements $m_{ij}$ against the number of photons reaching the detector.
This analysis was performed on only one SAWFS, at 775~nm.
The results are summarized on figure~\ref{fig:noise_propagation_chart}.
Figure~\ref{fig:noise_propagation_chart} shows photon and readout noise theoretical limits, computed from Fourier simulations, as well as photon noise corrected by a $\sqrt{2}$ factor, as is appropriate for the multiplicative noise of the Andor EMCCDs. It should be noted that quantum efficiency has not been accounted for and that the x-axis should be shifted appropriately depending on the sensor used.
Experimental points are computed, at each flux level, as the statistical average of the measurement error over 1000~consecutive frames.
The RMS error over all 21 baselines are shown, as well as the least and most noisy baselines.

We note first that the trend below $2\ 10^4$ ph/frame appropriately follows the -1/2 power law of photon noise, and that the least-noisy baselines are marginally above the photon noise limit.
For all measurements, the spread in RMS error between least and most noise baselines is of a factor $\approx$3, which could possibly be explained by illumination variations in the pupil plane; greater vibrations across long baselines; or slight misalignments of certain NRM holes overlapping with spider arms.

The performance bottoms out at 0.1 rad RMS averaged over baselines for illuminations beyond $2\ 10^4$~ph/frame. The best baseline however maintains a power -1/2 decrease up to almost $10^5$~ph/frame.
We suppose that floor in the error budget corresponds to instrumental vibrations and turbulence internal to the optical bench during the measurements: both as global aberrations, i.e. variations in the $\phi_i - \phi_j$, and as speckle noise from small-scale turbulence.

				\subsection{Noise constraints of the differential dewrapper}

We have shown in section~\ref{sec:reduction:dual} that for a capture range beyond $\pm \lambda^0 / 2$ and up to $\pm 2.36$~\textmu m with the dual band SAWFS, we need to measure $\delta d_{ij}$ to better than 50~nm of error.
This noise constraint is actually very strong -- and probably is, at this stage, one of the weakest points of the current choice of wavelengths.
This can be optimized in the future, possibly using more than just 2 channels.
This would enable finding the group delay using least-square methods, with more robustness to noise than the basic 2-channel method that we're implementing here.

A 50~nm error at 2-$\sigma$ extrapolates to 0.15~rad RMS per baseline per camera.
Looking at the experimental measurements shown on figure~\ref{fig:noise_propagation_chart}, this relates to respectively: $8\ 10^3$ ph/frame for the mean baseline overall, $3\ 10^3$ ph/frame for the mean baseline overall, and $2\ 10^4$ ph/frame for the noisiest baseline!
This last value would relate to a limiting magnitude at about $R=+4.4$ for 7~ms exposures (as used in section~\ref{sec:sky_ol}), considering the throughput of the VAMPIRES instrument.

However, the photon limits listed above stand as flexible bounds.
We interpret from figure~\ref{fig:noise_propagation_chart} that the RMSE floor at high flux comes from local turbulence or vibrations.
In this case, the source is actually physical and is correlated between cameras, thus canceling out when computing the $\delta d_{ij}$.
The eventual unwrapped OPD measurement $D_{ij}$ suffers from these perturbation sources, yet they do not compromise the success of the dewrapping.

Furthermore, we could also relax the 2-$\sigma$ separation constraint decided before. Slightly higher noise would induce a higher fraction of misses by the differential dewrapper.
However, many such misses can be detected, and the frame discarded, e.g. by computing and checking the integrity of closure phases on each frame before sending the resulting command to the DM.
The temporal structure of petal-locking, and the intended use of the SAWFS as a ``rectifier'' control over the high-speed AO loop, should permit a slow cadence, and thus it shouldn't be critical to discard identified dubious frames.
Moreover, the noise propagation through the linear reconstructor (eq.~\ref{eq:reconstruction}) is such that a 1~$\lambda$ error on any single baseline propagates into at worst 0.2~$\lambda$ peak-to-valley error on the reconstructed petal map.
This means that as long as dewrapper errors are occasional occurences, this additional error budget could be easily absorbed by the main AO loop, or by the SAWFS control law itself.

We will get back to differential dewrapper failure rate in section~\ref{sec:sky_ol:results_noise}, as we explore the statistics of the OPD difference $\delta d_{ij}$ in an on-sky dataset.

			\section{Open-loop measurements on-sky}
			\label{sec:sky_ol}
				
				\subsection{Observation conditions}
				\label{sec:sky_ol:conditions}

We performed on-sky engineering for the dual-band SAWFS on Subaru/SCExAO on the night of 2022/03/24, on Delta Virginis (R = +1.86, H = -1.05).
The VAMPIRES cameras were configured using the 7 hole NRM, with the narrowband configuration (675-25~nm and 775-25~nm filters respectively, after a 700~nm dichroïc), i.e. identically to the setup used during daytime tests detailed in the previous sections.
The cameras operate at 142.8~Hz in CCD frame transfer mode, with a 7~ms exposure time.
Camera shutters are synchronized by means of a hardware trigger.
In parallel, the fast imaging camera of SCExAO was acquiring focal plane images in H-band (1630~nm) with an exposure time of 400~\textmu s.
Seeing conditions were sub-median and unstable, with a baseline value of 0.8 - 1.0'' but with frequent bursts in the 1.5 - 2.0'' range, causing both AO loops to struggle to maintain consistent correction or even diffraction limited imaging in H-band.
Unfortunately, due to the AO correction instability, only a fraction of the 20 minutes of data acquired with the SAWFS could be analyzed at this point.
We present in this section some results of this analysis, focusing on a 11~second portion starting on 2022/03/24 11:30:56.7 AM UTC, comprised of 1500 frames for the SAWFS.
	
A few key results are analyzed through the following subsections: an a posteriori noise regime determination, temporal analysis of reconstructed petal piston statistics, and finally tentative H-band PSF reconstruction.

				\subsection{Results: noise regime for the dewrapper} 
				\label{sec:sky_ol:results_noise}

\begin{figure}[b] 
	\centering
	\includegraphics[width=.99\textwidth]{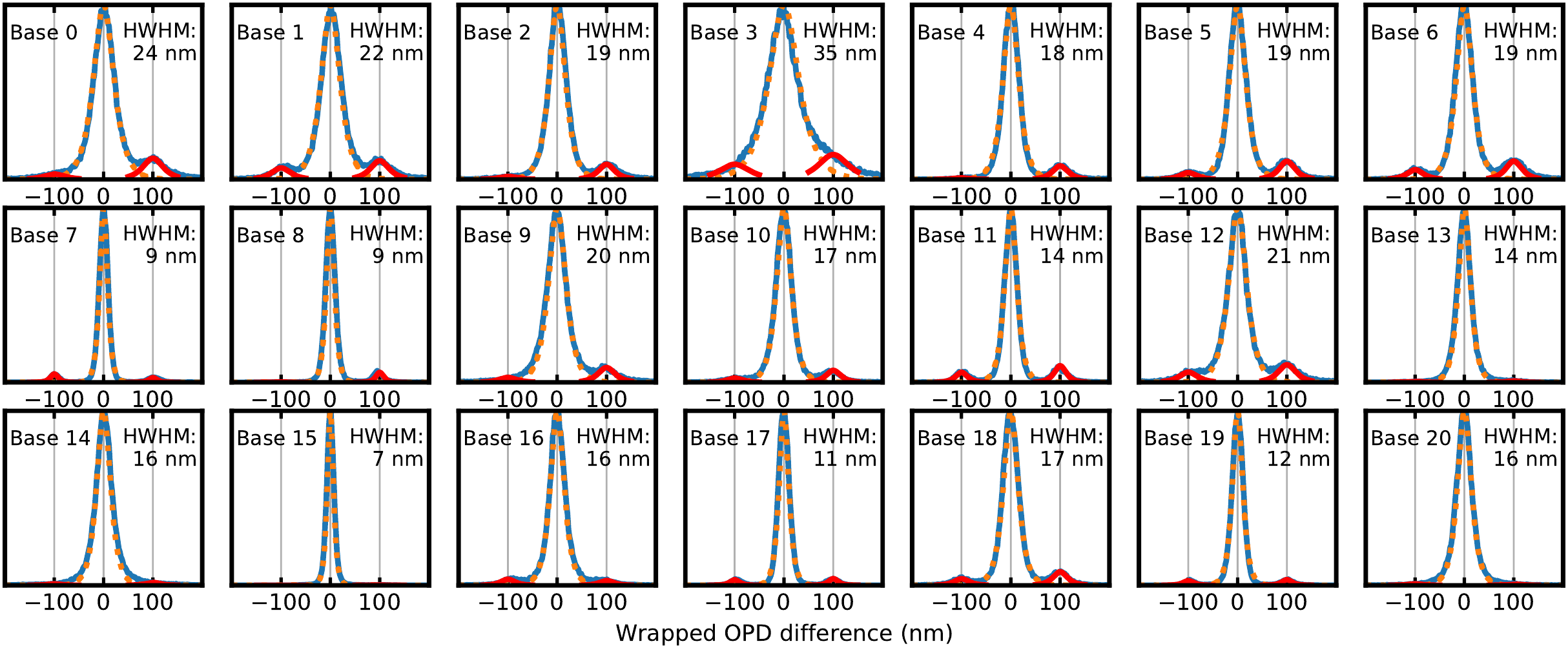}\\[.5em]
	\caption{%
		In blue, histograms of the differential OPD distribution \emph{prior to differential dewrapping} on each of the 21 SAM baselines during the on-sky sequence.
		Histograms are normalized to the height of the main peak.
		Histogram displays are trimmed to $\pm$200~nm, thus not showing the sidelobes at $\pm$675~nm.
		The half-width-at-half-maximum of the central peak is shown on each insert, ranging from 7 (baseline 15) to 35~nm (baseline 3).
		Dashed orange lines show a hyperbolic distribution with the same width as the central peak.
		Solid green lines show hyperbolic distributions with the same width as the central peak and same height as the histogram at the $\pm100$~nm sidelobes.
		Baselines 13, 14, 15 and 20 show no sidelobes as these baselines are internal to a single petal.
	}
	\label{fig:onsky_apost_histograms}
\end{figure}

The key requirement on noise for the differential dewrapper that arose throughout sections~\ref{sec:reduction:dual} and~\ref{sec:noise} can be further explored with the baseline phase statistics during our on-sky sequence, so as to assess a posteriori what the success rate of our dewrapping algorithm would have been.
We show on figure~\ref{fig:onsky_apost_histograms} the distribution of the differential OPD $\delta d_{ij}$ prior to differential unwrapping, on each of the 21 baselines of the NRM.

The central component of each $\delta d_{ij}$ histogram is for each baseline higher than the $\pm$100 sidelobes by at least an order of magnitude, allowing us to adjust a distribution for the structure of the peaks.
Empirically, we find the peak is well described by an hyperbolic secant distribution (pdf($x$) $\propto \mathrm{sech}\left(\frac{\pi}{2}\frac{x}{\sigma}\right)$) -- with this distribution fitting significantly better than either a Gaussian or Lorentzian.

Having an analytical distribution fitted to the $\delta d_{ij}$ allows us analyze noise statistics and to compute the failure rate of the differential dewrapper for the amount of $\delta d_{ij}$ spread observed on each baseline.
With HWHMs ranging from 7 to 35 nm, the differential dewrapper failure rate --i.e. the fraction of the surface area of each peak falling beyond $\pm50$~nm from its center-- ranges from $1.1\ 10^{-4}$ up to 0.19.
Interestingly, assuming independent processes on all 21 baselines, the dewrapper success rate on all baselines for a given frame would only be 1.3\%.
Even though a proper metric is lacking at this stage of the study, we can assess from petal values and PSF reconstructions (sections~\ref{sec:sky_ol:results_temporal} and~\ref{sec:sky_ol:results_psf}) that the success rate is significantly higher than 50\%, thus hinting that high noise occurences are highly correlated across baselines.
A likely candidate for such a process would be a large amount of high-order wavefront residuals from the AO, resulting in large speckle noise over all NRM holes.

Table~\ref{tab:failure_rates} briefly summarizes the relationship between peak standard deviation, HWHM, dewrapper failure rate, and per-camera noise across a general range of values, based on the analytical distribution fitted on the histograms of figure~\ref{sec:sky_ol:results_noise}.

\begin{table}[h!] 
	\centering
	\caption{%
		Computed single-baseline differential dewrapper failure rates, i.e. probabilistic occurrences of  $|\delta d_{ij}| > 50$~nm, based on an empirical hyperbolic distribution of the differential OPD $\delta d$.
		Table shows half-width-half-maximum and standard deviation (std) of the $\delta d$ distribution, as well as the noise in rad RMS on each baseline per camera that maps to the resulting $\delta d$.
	}
	\label{tab:failure_rates}
	\vspace*{0.5em}
	\resizebox{\textwidth}{!}{%
		\begin{tabular}{|l|c|c|c|c|c|c|c|c|c|c|c|}
			\hline
			HWHM of $\delta d_{ij}$ (nm)& 5 & 10 & 15 & 20 & 25 & 30 & 35 & 40 & 50 & 60 & 70\\
			\hline
			std of $\delta d_{ij}$ (nm)& 6.0 & 11.9 & 17.9 & 23.9 & 29.8 & 35.8 & 41.8 & 47.7 & 59.7 & 71.6 & 83.5\\
			\hline
			Failure rate (\%) & $2.4\ 10^{-4}$ & 0.17 & 1.6 & 4.7 & 9.1 & 14 & 19 & 24 & 33 & 41 & 47\\
			\hline
			Per camera (rad RMS) & $3.6\ 10^{-2}$ & $7.3\ 10^{-2}$ & 0.11 & 0.15 & 0.18 & 0.22 & 0.26 & 0.29 & 0.36 & 0.44 & 0.51\\
			\hline
		\end{tabular}
	} 
\end{table}
	
Now, as an exercise, we assume an identical photon flux on both cameras, and thus an identical readout/photon noise variance in radians on each camera, and that the histogram spreads observed on figure~\ref{fig:onsky_apost_histograms} are the direct product of only measurement noise phenomena such as discussed in section~\ref{sec:noise}.
For the experimental data, we note that 7~nm HWHM then maps to 0.051~rad RMS per camera, and 35~nm HWHM to 0.255~rad RMS.
Mapping these values back to the experimental results shown on figure~\ref{fig:noise_propagation_chart}, we notice that (1) the spread between the least- and most-noisy baselines is slightly above the experimental half-decade; and that (2) such values would place this experiment in the $10^4$-$2\ 10^4$ photons/frame range.
Zero-point calibrations for VAMPIRES place this dataset at $6\ 10^5$ ph/frame using R=+1.86 for Delta Vir.
Placed within figure~\ref{fig:noise_propagation_chart}, this dataset sit further to the right of the calibration source measurement point, atop a noise floor.
The latter could be caused by any combination of seeing, temporally varying AO residuals, or speckle noise, but in all cases, resulting in an RMSE way above the photon noise floor for such illumination.

				\subsection{Temporal analysis}
				\label{sec:sky_ol:results_temporal}
	
\begin{figure}[ht] 
	\centering
	\includegraphics[width=.99\textwidth]{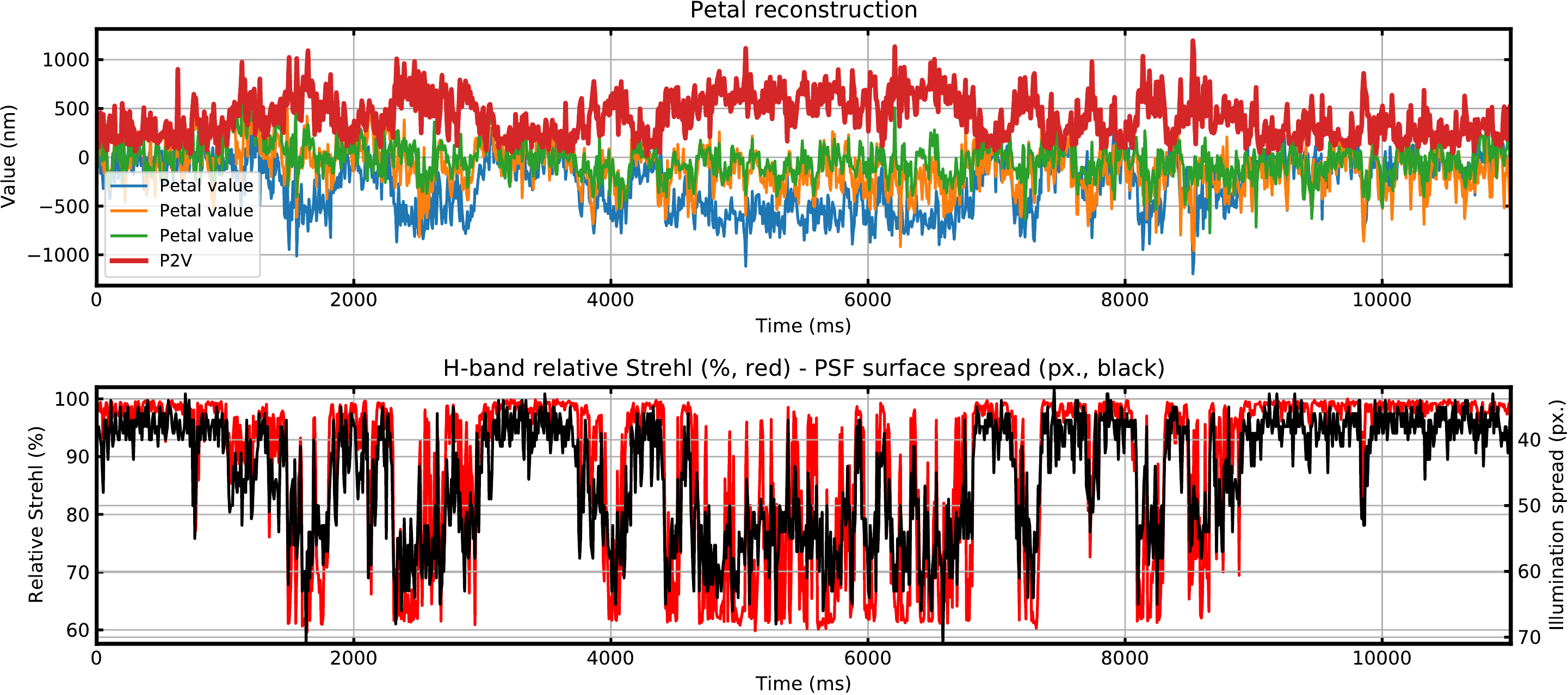}
	\caption{%
		Temporal series of the relative H-band Strehl ratio against the modal values of the 3 petal piston modes.
		Top: values of the 3 petal modes and the total peak-to-valley petaling amplitude reconstructed from the SAWFS measurements.
		Bottom: In red, Strehl ratio of the H-band PSF, relative to the maximum in this set. In black: the number of PSF detector pixels exceeding the value of 3000 ADU.
	}
	\label{fig:onsky_timeseries}
\end{figure}

In order to assess whether the open-loop measurement performed with the 2 band SAWFS on sky were successful, we analyze the reconstruction of the 3 petal modes from both cameras after differential dewrapping, and compare this value to the infrared PSF, used as an objective measure of wavefront quality.
In terms of peak-to-valley amplitude, the reconstructed petal wavefront exceeds $\lambda^0 / 2$ for 51\% of the samples. The dewrapper reconstructed baseline values in excess of $\lambda^0 / 2$ for 21\% of baseline measurements.

On figure~\ref{fig:onsky_timeseries}, we show on a time-base synchronized with the SAWFS an empirical Strehl and a surface-spread measurement, both from an H-band PSF.
The empirical Strehl is measured as the ratio between the PSF maximum to the maximum of the entire dataset.
It should be noted that this measure stays for extended periods in the 95-100\% range with very little fluctuations.
This is due to detector saturation during this sequence and therefore a loss of information on the true PSF maximum, rather than actual PSF stability.
The PSF spread is measured as the number of detector pixels that exceed a given illumination threshold, and varies between 32 and 71 pixels.
This spread metric is less sensitive to detector saturation and we observe even better correlation with the peak-to-valley amplitude of the petaling modes: we find a correlation coefficient of 0.74 between the petal peak-to-valley and the PSF spread, which tends to indicate meaningful measurements from the SAWFS.

				\subsection{PSF reconstruction}
				\label{sec:sky_ol:results_psf}
				
\begin{figure}[ht] 
	\centering
	\includegraphics[width=.99\textwidth]{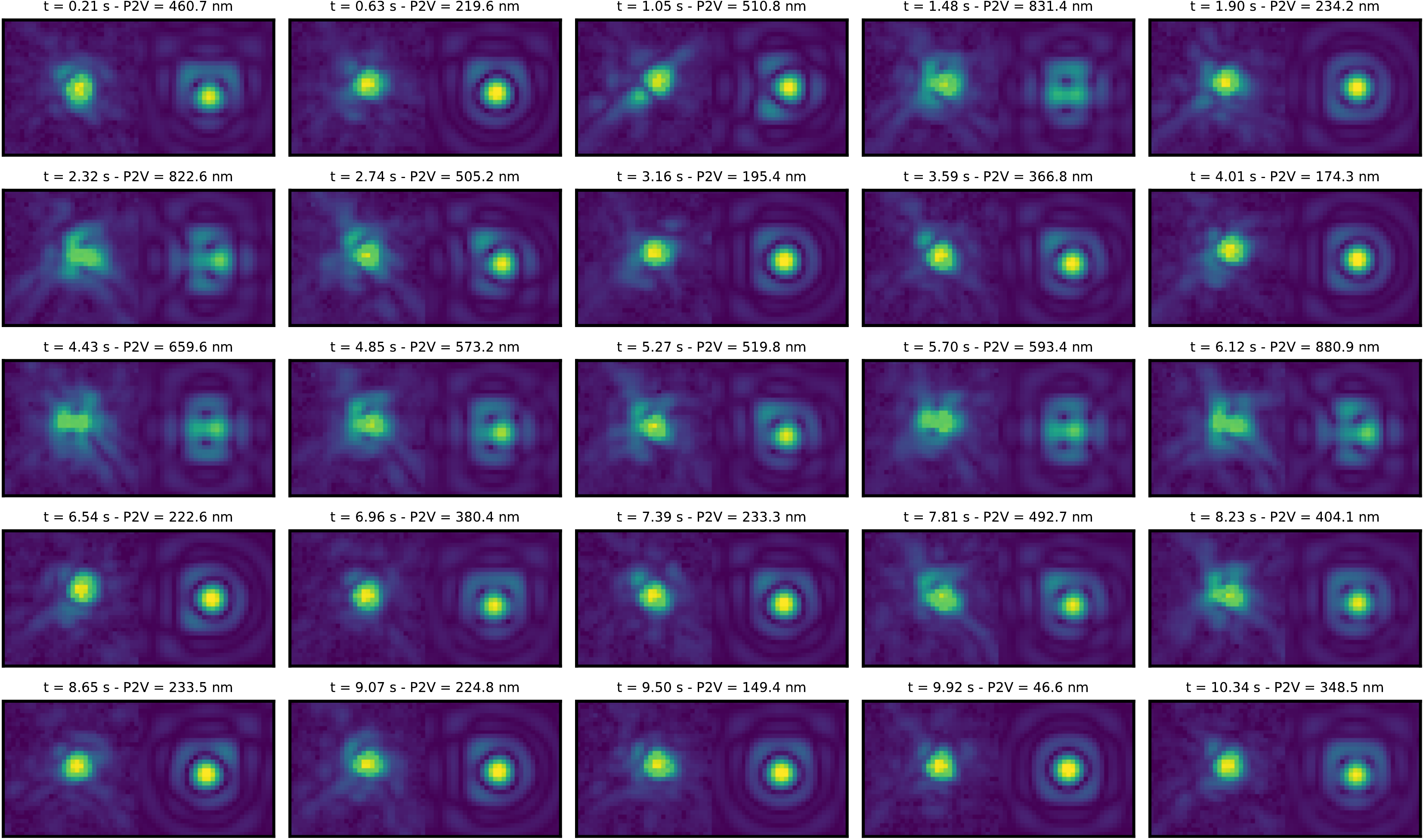}
	\caption{%
		Samples of H-band PSF reconstruction from SAWFS 3-mode petal data.
		For each insert, the left half shows the on-sky 1\,630~nm PSF, and the right half shows the petal mode reconstruction thereof.
	}
	\label{fig:onsky_chuck_recon}
\end{figure}

{
The correlation observed in section~\ref{sec:sky_ol:results_temporal} is however not proof that the SAWFS is performing a proper measurement of the petal modes in the wavefront - merely that it measures something that is correlated with PSF quality.

As to confirm that the SAWFS is performing proper measurements, we performed a direct reconstruction of the H-band PSF from the values measured by the SAWFS, using a direct approach: only far field propagation is simulated, through the Subaru pupil mask, aberrated with only the 3 free petal modes.
We show on figure~\ref{fig:onsky_chuck_recon} some sample results of this reconstruction.

We observe that the main distortion of the PSF core and the area of the first Airy ring are broadly reproduced by the reconstruction, showing that the process is somewhat consistent, and that we successfully reconstruct to better than the 1 wavelength ambiguity pitfall.
Some PSFs, e.g. at 5.70~s, could be excellent matches pending we would additionally simulate the effect of the partially corrected seeing in this $\approx$20-40\% Strehl ratio regime.

Obviously, the seeing conditions during this engineering run would not permit any sort of high-quality matching between the reconstructed PSF and the actual recorded PSF. Also, with only 3 petal degrees of freedom, the simulated PSF is restricted to a limited number of features.
The reconstruction could be improved already by adding global or petal tip-tilt into the reconstruction, which could be partially doable with the 7 hole NRM we are using.
}

			\section{CONCLUSION AND PERSPECTIVES}
			\label{sec:conclusion}
{
The impact of petaling on 8~m-class telescopes, and its projected aggravation on upcoming Extremely Large Telescopes, is likely to make the design and use of auxiliary WFSs a cornerstone of AO-assisted imaging, in particular for high contrast techniques, so as to circumvent the limitations of the PyWFS with the sensing of discontinuous aberrations at the spider arms.

We have shown that the use of non-redundant masking can enable proper petaling reconstruction, and provide an effective side channel to maintain the PyWFS of SCExAO within its unambiguous capture range in regard to petaling aberrations.
Once an adequate fusion control scheme between the main XAO loop and the SAWFS are implemented, we should be able to consistently improve scientific throughput during nights of prevalent low-wind effect.

We have shown that the use of two parallel spectral channels on synchronized detectors permits the recovery of discontinuous aberrations up to several wavelengths of dynamic range. This enables us to use a dual-channel SAWFS in visible light, a design compatible with scientific observations in the near-infrared -- which account for most of the high-contrast scientific observations performed on Subaru/SCExAO.

It has arisen during this work that unwrapping the phase beyond the half-wavelength capture range is actually a highly noise sensitive process with only two wavelengths.
Work is in progress to improve the unwrapping algorithm for two wavelengths. And in parallel, a redesign of VAMPIRES is currenty being studied that would include successive dichroics of neighboring cutoffs. This would allow multispectral imaging with four or more channels imaged on the same detector, and using NRMs would permit multi-band wavefront sensing, either in parallel with near-infrared science or simultaneously with visible light interferometry using that same NRM.
With more spectral channels, we would be able to implement more robust group delay retrieval techniques that are already well established in interferometry.

On-sky engineering has shown that we obtain a basic yet consistent reconstruction of the LWE-distorted PSF. We hope to be able to confirm these results in better seeing and during a night where there is no physical LWE, so as to obtain precise quantitative results over petaling aberrations introduced artificially.
}%

			\acknowledgments
			\label{sec:acknowledgements}
{
This work is based on data collected at Subaru Telescope, which is operated by the National Astronomical Observatory of Japan.
The authors wish to recognize and acknowledge the very significant cultural role and reverence that the summit of Maunakea has always had within the Hawaiian community.
We are most fortunate to have the opportunity to conduct observations from this mountain.
The authors also wish to acknowledge the critical importance of the current and recent Subaru Observatory daycrew, technicians, telescope operators, computer support, and office staff employees.
Their expertise, ingenuity, and dedication is indispensable to the continued successful operation of these observatories.
The development of SCExAO was supported by the Japan Society for the Promotion of Science (Grant-in-Aid for Research \#23340051, \#26220704, \#23103002, \#19H00703 \& \#19H00695), the Astrobiology Center of the National Institutes of Natural Sciences, Japan, the Mt Cuba Foundation and the director's contingency fund at Subaru Telescope. VD and NS acknowledge support from NASA (Grant \#80NSSC19K0336); KA acknowledges support from the Heising-Simons foundation.
}%

\bibliographystyle{spiebib} 
\bibliography{doctorat} 

\appendix
\renewcommand{\theequation}{\thesection.\arabic{equation}}
\numberwithin{equation}{section}
\renewcommand{\thefigure}{\thesection.\arabic{equation}}
\numberwithin{figure}{section}

\end{document}